\documentclass[aps,prl,reprint,superscriptaddress,amsmath,amssymb]{revtex4-2}
\usepackage{graphicx}
\usepackage{bm}
\usepackage{physics}

\usepackage{hyperref}
\usepackage{placeins}

\usepackage{fancyhdr}
\begin{document}

\title{Quadrature-Symmetric PulsePol for Robust Quantum Control Beyond the Ideal Pulse Approximation}
\author{Mayur Jhamnani}
\affiliation{Center for Quantum and Topological Systems, New York University Abu Dhabi, PO Box 129188, Abu Dhabi, United Arab Emirates}
\author{Venkata SubbaRao Redrouthu}
\affiliation{Center for Quantum and Topological Systems, New York University Abu Dhabi, PO Box 129188, Abu Dhabi, United Arab Emirates}
\author{José P. Carvalho}
\affiliation{Interdisciplinary Nanoscience Center (iNANO) and Department of Chemistry, Aarhus University, Gustav Wieds Vej 14, DK-8000 Aarhus C, Denmark}
\author{Ethan Feldman}
\affiliation{Center for Quantum and Topological Systems, New York University Abu Dhabi, PO Box 129188, Abu Dhabi, United Arab Emirates}
\author{Anders B. Nielsen}
\affiliation{Interdisciplinary Nanoscience Center (iNANO) and Department of Chemistry, Aarhus University, Gustav Wieds Vej 14, DK-8000 Aarhus C, Denmark}
\author{Phani Kumar}
\affiliation{Indian Institute of Science Education and Research, Bhopal, 462066, India}
\author{Niels Chr. Nielsen}
\affiliation{Interdisciplinary Nanoscience Center (iNANO) and Department of Chemistry, Aarhus University, Gustav Wieds Vej 14, DK-8000 Aarhus C, Denmark}
\author{P. K. Madhu}
\affiliation{Tata Institute of Fundamental Research Hyderabad, Hyderabad, 500046, India}
\author{Asif Equbal}
\email{asif@nyu.edu}
\affiliation{Center for Quantum and Topological Systems, New York University Abu Dhabi, PO Box 129188, Abu Dhabi, United Arab Emirates}
\affiliation{Department of Chemistry, New York University Abu Dhabi, PO Box 129188, Abu Dhabi, United Arab Emirates}

\begin{abstract}
PulsePol is an elegantly designed pulse-sequence-based quantum control scheme that enables polarization transfer between electron and nuclear spins, for example, in nitrogen-vacancy (NV) centers. However, previous analyses of PulsePol assumed very strong, close to ideal, instantaneous microwave pulses, which is rarely achievable as one goes to higher magnetic fields. We revisit the PulsePol scheme under finite-pulse constraints and show that its performance significantly degrades because of finite-pulse effects. Using bimodal Floquet theory, we identify the symmetry-breaking mechanism responsible for this deteriorating fidelity. By phase adjustment, we reestablish the proper symmetry of the interaction-frame spin Hamiltonian---leading to a sequence called Q-PulsePol, where 'Q' reflects the restored quadrature symmetry. Our results demonstrate robustness to finite-pulse effects and improved polarization transfer efficiency, establishing Q-PulsePol as a practical and reliable scheme for bulk hyperpolarization of nuclear spins in solid-state using a single-mode (zero-quantum or double-quantum) transfer. This work bridges idealized quantum control with realistic pulse engineering, establishing design rules for spin-based quantum-control protocols.
\end{abstract}

\maketitle

\thispagestyle{fancy}
\footnotetext[1]{Corresponding author: \texttt{asif@nyu.edu}}

\section{Introduction} \label{1}
Solid-state systems hosting electron spins, such as persistent radicals and nitrogen-vacancy (NV) centers in diamond, are prominent platforms for quantum information processing, communication, and sensing. In such systems, electron spins can be optically initialized to a polarization of above 90\% and can exhibit coherence times of tens of microseconds even at room temperature \cite{degen2017quantum, wrachtrup2010defect, west2019gate, bradley2019ten, bradley2022robust, abobeih2022fault, qiu2023enhancing, qiu2023optical, yamauchi2024toward, le2025coherent}. However, to extend quantum lifetimes further, attention has turned to proximal nuclear spins, which maintain coherence times of seconds to minutes owing to their smaller magnetic moments and, as a result, weak coupling to environmental noise. A key challenge, however, is the extremely low thermal polarization of nuclear spins ($\sim10^{-6}$ at room temperature), limiting both readout signals in NMR/MRI and the purity of initial states in quantum registers. Dynamic Nuclear Polarization (DNP) overcomes this by transferring polarization from hyperpolarized electron spins to coupled nuclear spins via the continuous irradiation of resonant microwaves ($\mu$w) \cite{carver1953polarization, overhauser1953paramagnetic, abragam1955overhauser, jeffries1960dynamic, abragam1961principles, hwang1967phenomenological, borghini1968spin, abragam1978principles, ernst1990principles, ardenkjaer2003increase, maly2008dynamic, glenn2018high, equbal2019cross, wolfowicz2021quantum, wili2024observation, javed2025dynamic}.

\begin{figure*}[t]
\centering
\includegraphics[width=0.9\textwidth]{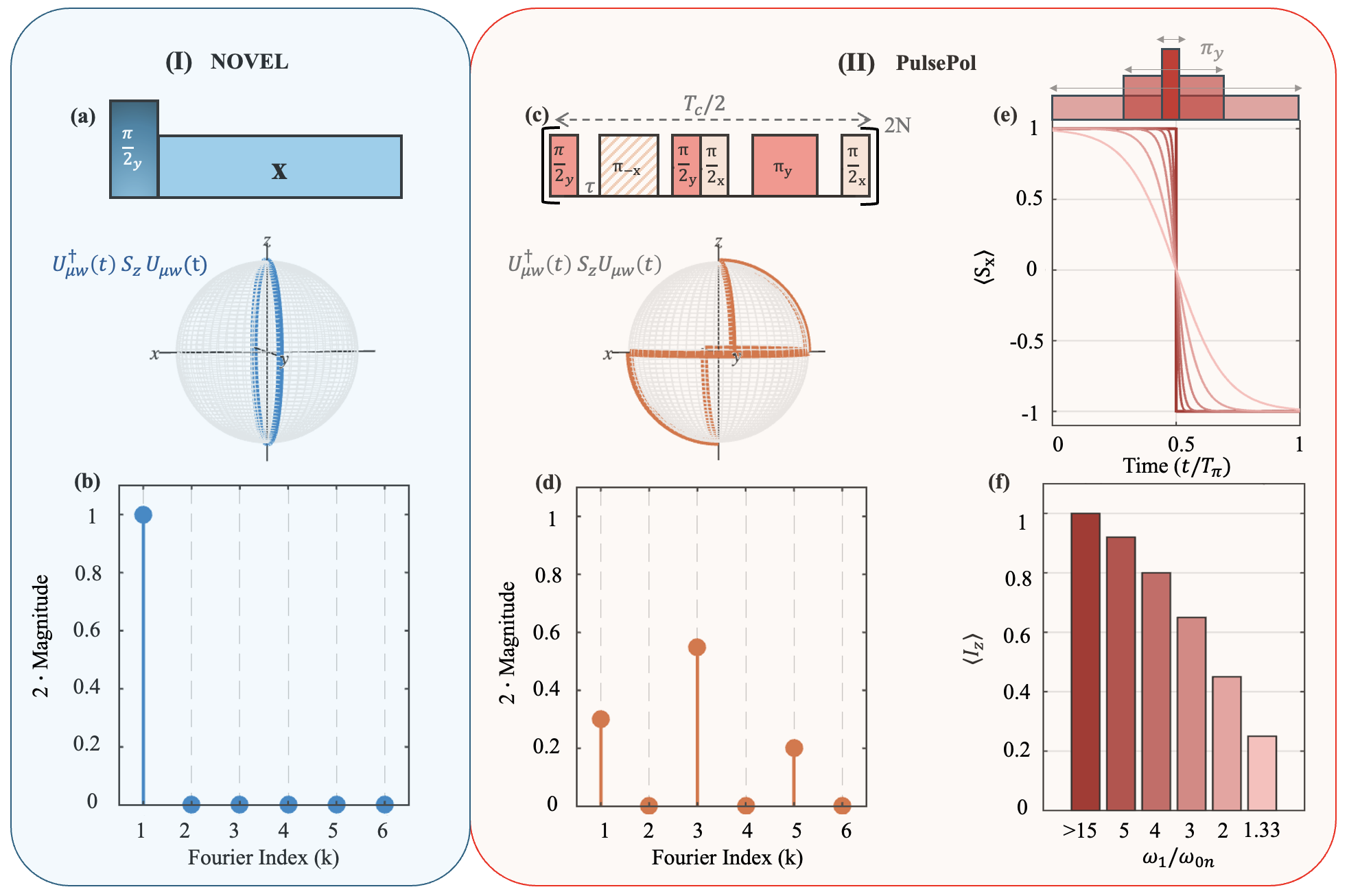}
\caption{\textbf{The Problem in PulsePol and Need for Hamiltonian Engineering.} Comparison of NOVEL and PulsePol pulse-sequences and their performance under finite-pulse conditions. (I) NOVEL: (a) pulse schematic and $e^-$ spin trajectory in the interaction frame, showing (b) a single dominant Fourier mode. (II) PulsePol: (c) pulse schematic and $e^-$ spin trajectory (see Eq. (5)), which is cyclic only over an even number of cycles. The free-precession delay $\tau$ is set by the DNP resonance condition $\omega_{0n} = k\omega_c$, which fixes the cycle time $T_c$ ($\omega_c = 2\pi/T_c$). (d) Fourier spectrum (see Eq. (6)) revealing multiple active resonance conditions with dominance at $k=3$. (e) $e^-$-spin inversion profile of a $\pi$-pulse along $S_y$, illustrating how increasing pulse finiteness ($T_{\pi}$ denotes the duration of the longest $\pi$ pulse) progressively distorts the ideal square-wave trajectory (red), breaking the toggling-frame symmetries that PulsePol relies upon. (f) Consequently, the nuclear polarization $\langle I_z \rangle$ (plotted for short contact time $\sim10-20 \,\mu$s for $\omega_{0n}/(2\pi) \approx 15$ MHz) degrades rapidly with increasing pulse finiteness (gradient corresponding to (e) highlights arbitrarily how pulse-finiteness decreases PulsePol efficiency).}
\label{fig:fig1}
\end{figure*}

While continuous-wave DNP is standard at high magnetic fields ($B_0$), the recently introduced pulsed DNP offers coherent, selective quantum control at \textit{seemingly} lower average $\mu$w power (nutation frequency), mitigating sample heating \cite{equbal2019pulse, camenisch2024pulsed}. The Nuclear Orientation Via Electron spin Lock (NOVEL) was the first pulsed DNP sequence (Fig.~\ref{fig:fig1}(a)) \cite{can2015time}. However, NOVEL relies on the Hartmann-Hahn matching \cite{hartmann1962nuclear} of the electron nutation frequency $\omega_1$ to the nuclear Larmor frequency $\omega_{0n}$ and as a result requires very high $\mu$w powers at relatively high $B_0$. This has motivated considerable attention to the development of advanced multiple-pulse DNP experiments aimed at improving DNP efficiency for more general experimental conditions \cite{jain2017off, tan2019time, redrouthu2022efficient, wili2022designing, redrouthu2023dynamic, nielsen2024dynamic, javed2025magic, nielsen2025controlling, carvalho2025optimal, carvalho2026longitudinal}. Among these, with focus on NV-center-based DNP, PulsePol, introduced in 2018 \cite{schwartz2018robust}, emerged as a robust pulse-sequence: its symmetric phase pattern (Fig.~\ref{fig:fig1}(c)) generates polarization transfer under electron-spin resonance detuning \cite{levitt2008symmetry, tratzmiller2021parallel, sabba2022symmetry, espinos2024enhancing, munuera2025pulse, pal2025robust, blinder202513c}. While NOVEL is characterized by single-mode transfer (Fig.~\ref{fig:fig1}(b)), PulsePol involves multiple Fourier components at odd harmonics, and as a result, many resonance conditions (Fig.~\ref{fig:fig1}(d)).

Despite its strengths, PulsePol has only been analyzed considering very strong, instantaneous $\mu$w pulses \cite{schwartz2018robust}. In realistic settings, particularly for lower sample volumes or operation at high $B_0$ fields, $\mu$w power is limited, so pulses occupy a significantly finite fraction of the cycle time $T_c$. Lower $\mu$w power implies greater pulse length, elongating the pulse trajectory (Fig.~\ref{fig:fig1}(e)).

In this work, we demonstrate that PulsePol is not well suited in this bounded-control regime: finite pulse trajectories break the toggling-frame symmetries in PulsePol required for selective and robust DNP, activating competing zero-quantum (ZQ, flip-flop) and double-quantum (DQ, flip-flip) pathways that degrade DNP efficiency (Fig.~\ref{fig:fig1}(f)) and nullify net polarization diffusion to the bulk nuclear spins. We restore these symmetries by utilizing Quadrature-Symmetric PulsePol (Q-PulsePol), constructed via a minimal phase adjustment to the central inversion pulse in PulsePol, which we experimentally demonstrate to be robust to finite-pulse effects and highly efficient for bulk polarization transfer. Importantly, we develop a general framework for designing symmetric quantum control schemes for robust and efficient DNP.

To achieve this, we employ Floquet theory in the $\mu$w interaction frame. We first derive the effective interaction-frame Hamiltonian and the DQ/ZQ scaling factors, uncover the two fundamental Floquet symmetries (Quadrature and XY-Time-Reversal) that govern selective and efficient DNP, show explicitly how finite pulses break them in standard PulsePol, and demonstrate how the phase adjustment restores perfect symmetry for any pulse duration.

\section{Theoretical Background} \label{2}
Consider a coupled electron-nucleus system with the electron spin ($S$) in the rotating frame and the nuclear spin ($I$) in the lab frame:
\begin{equation}
\mathcal{H}(t) = \Omega S_z + A_z S_z I_z + A_x S_z I_x + \omega_{0n} I_z + \mathcal{H}_{\mu w}(t),
\end{equation}
where $\Omega$ is the $\mu$w offset, $A_z$ and $A_x$ are the secular and pseudo-secular hyperfine couplings respectively, and
\begin{equation}
\mathcal{H}_{\mu w}(t) = \omega_1(t) [S_x \cos\phi(t) + S_y \sin\phi(t)],
\end{equation}
where $\omega_1(t)$ is a step function (i.e., it is either zero or $\omega_1$). 

In the ideal pulse limit ($\omega_1 \gg \omega_{0n}$), the evolution under the internal Hamiltonian ($\mathcal{H}(t) - \mathcal{H}_{\mu w}(t)$) becomes negligible during the pulses and is effective only during the delay periods. However, for finite pulse limit ($\omega_1 \not\gg \omega_{0n}$), the contribution of the internal Hamiltonian cannot be neglected during pulses, and the full Hamiltonian $\mathcal{H}(t)$ governs the evolution of the system. The density matrix evolves as $\rho(t) = U(t)\rho(0)U(t)^\dagger$, where $U(t)$ is the time-evolution operator (propagator) corresponding to $\mathcal{H}(t)$. We employ bimodal Floquet theory to derive an effective time-independent Hamiltonian for $\mathcal{H}(t)$ \cite{waugh2007average, halder2026applications, scholz2007operator, equbal2016relative, ivanov2021floquet}.

For analytical convenience, we transform $\mathcal{H}(t)$ into the modulating interaction frame via the propagators
\begin{equation}
\begin{split}
U_{\mu w}(t) &= \mathcal{T} \exp \left( -i\int_{0}^{t} 
(\Omega S_{z} + \omega_{1}(t_{1})S_{x}) dt_{1} \right), \\
U_{0n}(t) &= e^{i\omega_{0n}I_{z}t}
\end{split}
\end{equation}
where $\mathcal{T}$ is the Dyson time-ordering operator. The interaction frame Hamiltonian is expressed as:
\begin{equation}
{\mathcal{\Tilde{H}}}(t) = {\Tilde{S}}_z(t) [A_z I_z + A_x (I_x \cos\omega_{0n}t + I_y \sin\omega_{0n}t)],
\end{equation}
with the back-propagated electron-spin trajectory (also referred to as \emph{spin-trajectories}) 
\begin{equation}
{\Tilde{S}}_z(t) = U_{\mu w}^\dagger(t) S_z U_{\mu w}(t).
\end{equation}
Under the microwave interaction, $S_z$ gets rotated into a combination of $S_x$, $S_y$, and $S_z$ at each moment. We refer to the resultant X- and Y-component of the spin-trajectories as $X(t)$ and $Y(t)$. Moreover, because the pulse-sequence is periodic with modulation or cycle frequency $\omega_c = 2\pi/T_c$, we use the Fourier series expansion:
\begin{equation}
{\Tilde{S}}_z(t) = \sum_{k=-\infty}^\infty \left[(a_x^{(k)} S_x + a_y^{(k)} S_y) + a_z^{(k)} S_z \right] e^{i k \omega_c t}.
\end{equation}
Substituting into ${\Tilde{S}}_z(t))$ and expressing the terms rotating with the nuclear Larmor frequency as a Fourier series, we can write a bimodal Floquet Hamiltonian:
\begin{equation}
{\mathcal{\Tilde{H}}}(t) = \sum_{n=-1}^{1} \sum_{k=-\infty}^{\infty} {\mathcal{\Tilde{H}}}^{(n,k)} e^{i n \omega_{0n} t} e^{i k \omega_c t}.
\end{equation}
Here, $n$ and $k$ denote the Fourier indices associated with the nuclear Larmor frequency and the microwave modulation frequency, respectively. The DNP-relevant Fourier component:
\begin{align}
{\mathcal{\Tilde{H}}}^{(\pm1,k)} &= \frac{A_x}{2} \bigl(a_x^{(k)} S_x + a_y^{(k)} S_y + a_z^{(k)} S_z\bigr) (I_{\pm}).
\end{align}
Resonant DNP occurs when $\omega_{0n} = \pm k\omega_c$, with efficiencies governed by the DQ and ZQ scaling factors,
\begin{equation}
\chi_{\rm DQ}^{(k)} = \sqrt{ \bigl[\operatorname{Re}(a_x^{(k)}) + \operatorname{Im}(a_y^{(k)})\bigr]^2 + \bigl[\operatorname{Im}(a_x^{(k)}) - \operatorname{Re}(a_y^{(k)})\bigr]^2 },
\end{equation}
\begin{equation}
\chi_{\rm ZQ}^{(k)} = \sqrt{ \bigl[\operatorname{Re}(a_x^{(k)}) - \operatorname{Im}(a_y^{(k)})\bigr]^2 + \bigl[\operatorname{Im}(a_x^{(k)}) + \operatorname{Re}(a_y^{(k)})\bigr]^2 }.
\end{equation}
These scaling factors are derived in SI-1. We note that similar expressions may be derived using single-spin-vector effective Hamiltonian theory \cite{shankar2017general, nielsen2019single, nielsen2022accurate}.

The relative magnitudes of $\chi_{DQ}$ and $\chi_{ZQ}$, and therefore whether a sequence produces pure or competing DNP pathways, are strictly controlled by fundamental symmetries in the Fourier components of the interaction-frame trajectory $\tilde{S}_z(t)$. We now analyze these Floquet symmetries in detail.

\subsection{Role of Quadrature Symmetry in the Floquet Hamiltonian} \label{3}

\begin{figure}[h]
    \centering
    \includegraphics[width=1.0\linewidth]{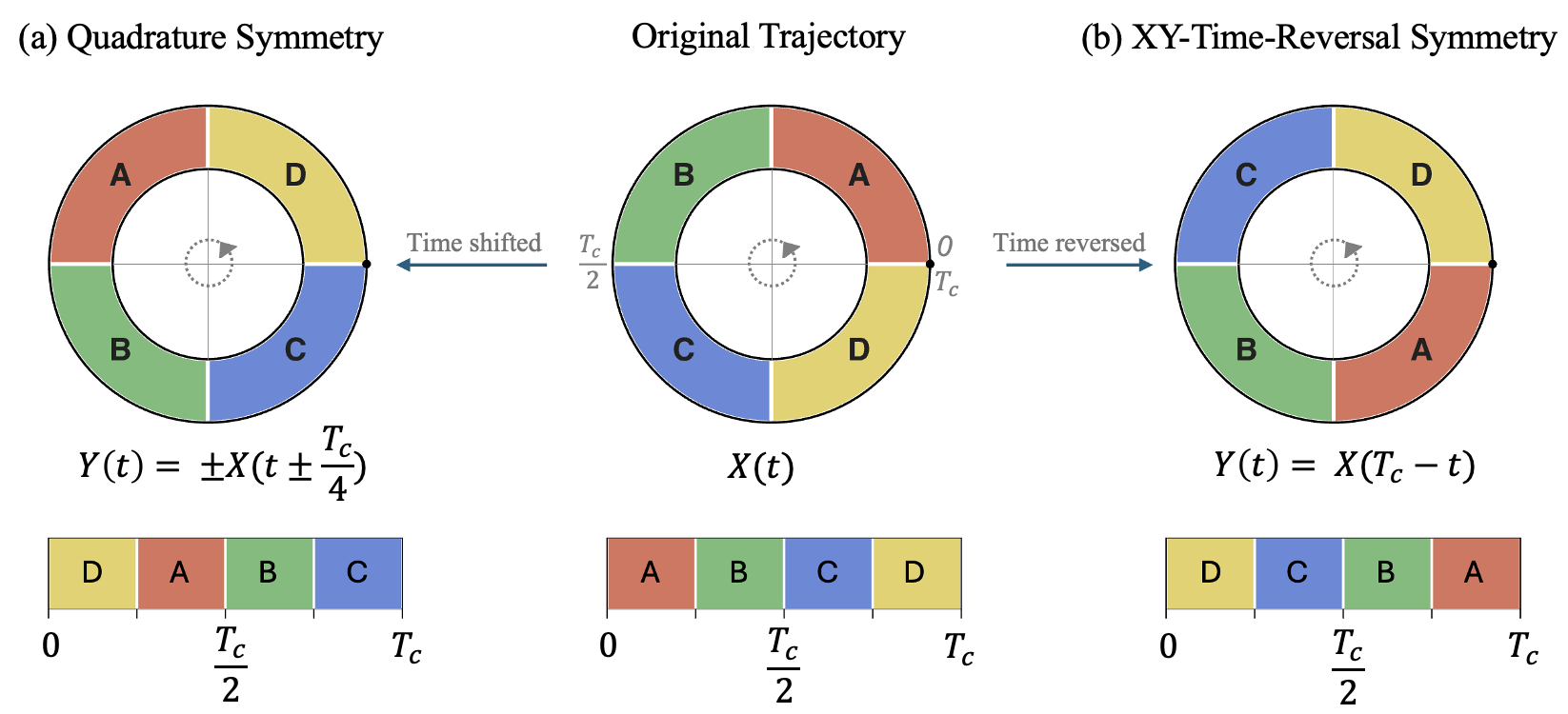}
    \caption{\textbf{Visualization of symmetry operations in the interaction-frame spin trajectories.} The central panel shows the original trajectory, or phase cycle, $X(t)$, composed of four segments (A--D) over one cycle period $T_c$. (a) \textit{Quadrature symmetry}: the Y-component follows from a quarter-cycle time shift of the X-component, $Y(t) = \pm X(t \pm T_c/4)$, corresponding to a cyclic permutation of the phase segments (ABCD $\rightarrow$ DABC). (b) \textit{XY time-reversal symmetry}: the Y-component is related to the X-component through time reversal, $Y(t) = X(T_c - t)$, resulting in a reversed ordering of segments (ABCD $\rightarrow$ DCBA). These symmetry constraints determine the characteristics of the Fourier coefficients and, consequently, the DNP scaling factors.}
    \label{fig:visual_symmetry}
\end{figure}

The key to realizing transition-selective DNP is the presence of specific symmetries in the interaction frame electron spin trajectory \({\Tilde{S}}_z(t)\). These Floquet symmetries control the relative magnitudes of \(\chi_{\rm DQ}^{(k)}\) and \(\chi_{\rm ZQ}^{(k)}\) through simple relations among the complex coefficients \(a_x^{(k)}\) and \(a_y^{(k)}\) (see Eq.~(6)), thereby enabling the design of sequences that activate one resonance pathway while completely suppressing the other.

\begin{figure*}[t!]
\centering
\includegraphics[width=\textwidth]{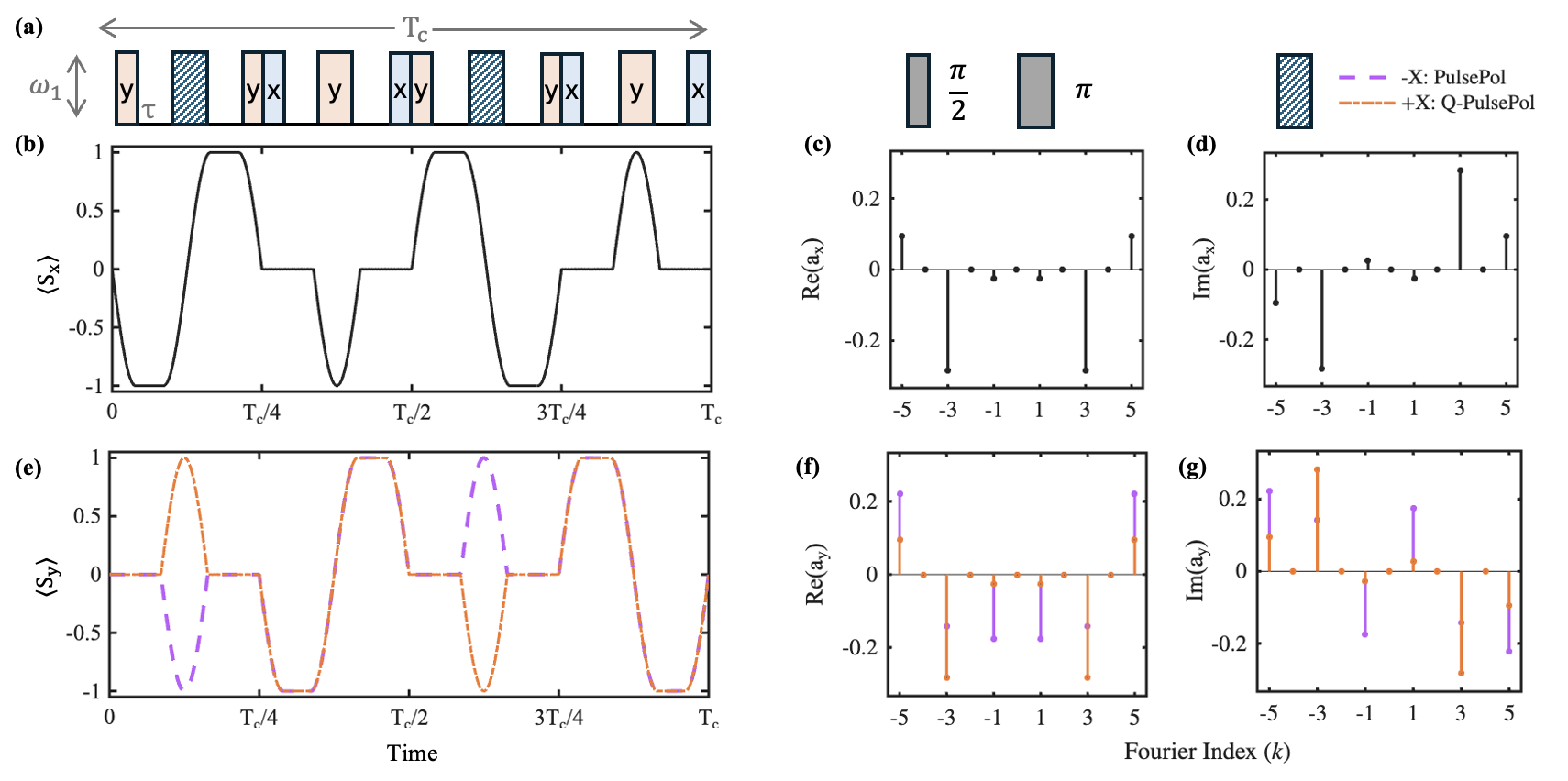}
\caption{\textbf{Fourier analysis of PulsePol and Q-PulsePol under finite pulses.} (a) Pulse-sequence schematics for standard PulsePol ($-$X central pulse, purple) and Q-PulsePol ($+$X central pulse, orange). A pulse phase of $360^\circ$ implies the completion of one $T_c$. (b,e) Interaction-frame $X(t)$ ($\langle S_x\rangle$) and $Y(t)$ ($\langle S_y\rangle$) components, respectively. Notice that the pulse-trajectories are finite, i.e., they are not delta functions and possess a slope. Critically, the central phase correction leaves $X(t)$ unchanged (black) but changes $Y(t)$ and helps in restoring the proper quadrature relationship between $X(t)$ and $Y(t)$. (c,d) Real and imaginary parts of the Fourier coefficients $a_x^{(k)}$, and (f,g) real and imaginary parts of $a_y^{(k)}$, confirming that Q-PulsePol satisfies the algebraic conditions of Eqs.\,(11) and (13), establishing pure DQ selectivity and recovery of high scaling factor, whereas standard PulsePol violates them both under finite pulses.}
\label{fig:finitepulse}
\end{figure*}

From Eqs.~(9) and (10), a pure DQ resonance (\(\chi_{\rm ZQ}^{(k)}=0\)) is obtained when
\begin{equation}
\operatorname{Re}(a_x^{(k)}) = \operatorname{Im}(a_y^{(k)}) \quad \text{and} \quad \operatorname{Im}(a_x^{(k)}) = -\operatorname{Re}(a_y^{(k)}).
\end{equation}
Conversely, a pure ZQ resonance (\(\chi_{\rm DQ}^{(k)}=0\)) occurs when
\begin{equation}
\operatorname{Re}(a_x^{(k)}) = -\operatorname{Im}(a_y^{(k)}) \quad \text{and} \quad \operatorname{Im}(a_x^{(k)}) = \operatorname{Re}(a_y^{(k)}).
\end{equation}
When a pulse-sequence satisfies either Eq. (11) or Eq. (12), we call it \emph{uni-modal}; as we select either one of DQ or ZQ-DNP transition, and never both simultaneously.

The algebraic conditions in Eqs. (11) an (12) are mathematically equivalent to \(a_y^{(k)} = i \, a_x^{(k)}\) (for pure DQ) or \(a_y^{(k)} = -i \, a_x^{(k)}\) (for pure ZQ). When the \(k\)-th harmonic is reconstructed from the Fourier expansion of \({\Tilde{S}}_z(t)\) (see SI-2), this relation implies that if the $X(t)$ varies as \(\cos(k \omega_c t + \phi)\), the corresponding $Y(t)$ varies as \(\mp\sin(k \omega_c t + \phi)\) depending on whether a DQ or a ZQ transition is chosen. Such a symmetry produces a circularly polarized modulation of the electron spin operators at frequency \(k\omega_c\) and selectively excites only one DNP transition (either DQ or ZQ), while the orthogonal pathway vanishes. This occurs only when the spin-trajectories in X and Y have the relation $Y(t) = \pm X(t \pm \frac{T_c}{4})$, or more simply, are in \emph{quadrature}, as any other relative phase would activate both DQ and ZQ pathways simultaneously. We therefore refer to this symmetry as \emph{Quadrature symmetry}. 

To visualize quadrature symmetry (see Fig. \ref{fig:visual_symmetry}a), consider dividing the interaction-frame spin trajectories into four equal quarters of the cycle time $T_c$, and labeling the trajectory segments in the X-component sequentially as $A$, $B$, $C$, $D$. Quadrature symmetry requires that the Y-component carries the same four segments but shifted by exactly one quarter cycle, giving, for example, $D$, $A$, $B$, $C$, such that $Y(t) = X(t - T_c/4)$ in this case.

Let a pure DQ resonance be selected via Quadrature symmetry, then, the scaling factor reaches its theoretical maximum when
\begin{equation}
\operatorname{Re}(a_x^{(k)}) = \operatorname{Im}(a_y^{(k)}) = -\operatorname{Im}(a_x^{(k)}) = \operatorname{Re}(a_y^{(k)}).
\end{equation}
This additional constraint is satisfied precisely when $Y(t)$ is the time-reversed version of the $X(t)$, \(Y(t) = X(T_c-t)\). We term this property \emph{XY-Time-Reversal symmetry}. 

A general representation of XY-Time-Reversal symmetry in the X and Y-components of the interaction frame trajectories is shown in Fig.~\ref{fig:visual_symmetry}b. Accordingly, if the X-component reads $A$, $B$, $C$, $D$, then XY-Time-Reversal symmetry requires the Y-component to read $D$, $C$, $B$, $A$, the same segments in reverse order. When both Quadrature and XY-Time-Reversal symmetries coexist, the reconstructed \(X\) and \(Y\) harmonics at all the resonant orders \(k\) have identical amplitudes and maintain a precise \(90^\circ\) phase difference.

\begin{figure}[h]
\centering
\includegraphics[width=0.965\linewidth]{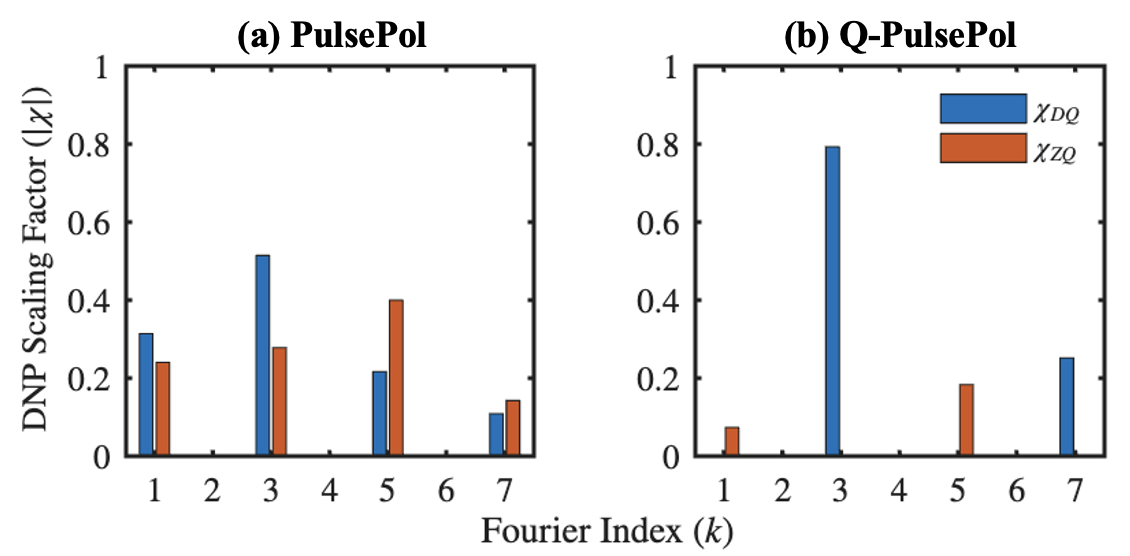}
\caption{\textbf{DNP Scaling Factors in the Finite Pulse Limit.} DQ and ZQ scaling factors $\chi_{\rm DQ}^{(k)}$ and $\chi_{\rm ZQ}^{(k)}$ as a function of Fourier index $k$ for (a) standard PulsePol and (b) Q-PulsePol, evaluated at the maximally finite pulse condition where pulse durations collectively occupy the full cycle time $T_c$ ($f = 1$, see Sec. \ref{4}). While standard PulsePol activates competing DQ and ZQ pathways at all harmonics, Q-PulsePol enforces quadrature symmetry and recovers pure uni-modal transfer. This demonstrates that the central phase correction restores 
selectivity even under the most demanding finite-pulse conditions.}
\label{fig:scaling}
\end{figure}

In the ideal-pulse limit, PulsePol (Fig.~\ref{fig:finitepulse}(a)) satisfies both Quadrature and XY-Time-Reversal symmetries because instantaneous rotations produce perfectly symmetric square-wave trajectories. However, at finite \(\omega_1\), both these symmetries are broken due to the central $-$X inversion pulse, as clearly visible in the spin trajectories in Fig.~\ref{fig:finitepulse}(b, e; see purple curve). Fourier analysis of the X- and Y-components (Fig.~\ref{fig:finitepulse}c, d, (f, g; see purple curve)) confirms that the algebraic conditions of Eqs.~(11) and (12) are violated, so that standard PulsePol no longer possesses Quadrature symmetry and therefore activates competing DQ and ZQ pathways (Fig.~\ref{fig:scaling}(a)).

To restore both symmetries in the finite-pulse limit, we change the phase of the central \(\pi\)-pulse from $-$X to $+$X. This minimal adjustment reestablishes Quadrature and XY-Time-Reversal symmetry for arbitrary pulse durations, as evident in Fig.~\ref{fig:finitepulse}(b, e; see orange curve). The corresponding Fourier analysis (Fig.~\ref{fig:finitepulse}c, d, (f, g; see orange curve)) confirms that the conditions of Eqs.~(11) and (13) are fully satisfied. Consequently, Q-PulsePol simultaneously achieves pure uni-modal DNP transfer and recovers a scaling factor significantly larger than PulsePol, as shown in Fig.~\ref{fig:scaling}(b).

The robustness of this symmetry restoration under experimental constraints is best quantified by examining how the dominant Fourier coefficients evolve with pulse finiteness, which we address in the following section.

\section{Pulse Finiteness} \label{4}

In general, DNP scaling factors \(\chi_{\rm DQ}^{(k)}\) and \(\chi_{\rm ZQ}^{(k)}\) at the dominant resonance condition, corresponding to Fourier index, $k$=3, are determined by the Fourier coefficients \(a_x^{(k)}\) and \(a_y^{(k)}\) through Eqs.~(9) and (10). Understanding how finite microwave pulses modify these coefficients is therefore essential.

\begin{figure}[h!]
\centering
\includegraphics[width=0.98\linewidth]{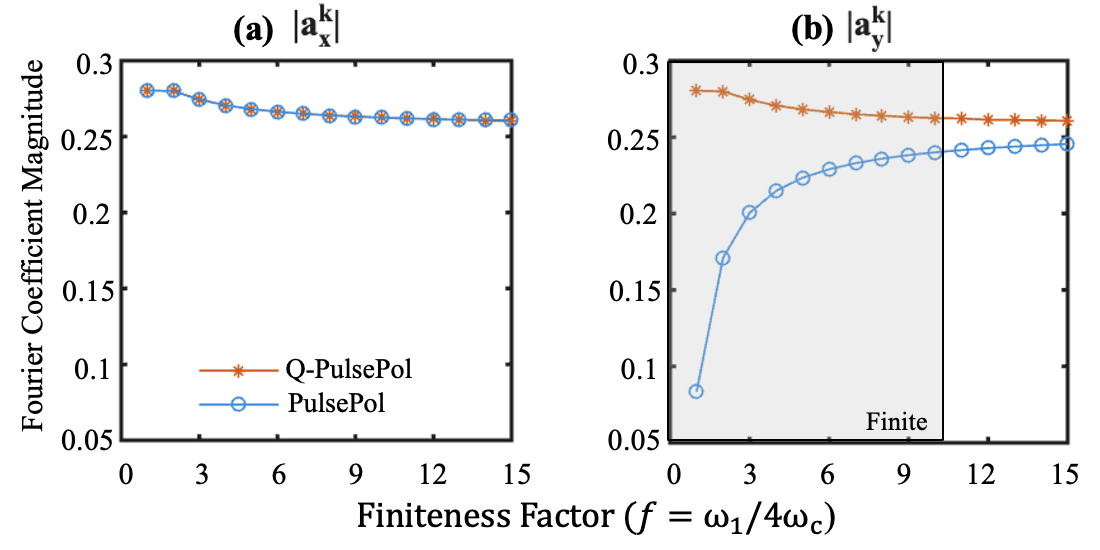}
\caption{\textbf{Effect of Pulse Finiteness on the Dominant Fourier Coefficients.} Magnitude of \(|a_x^{(k)}|\) (a) and \(|a_y^{(k)}|\) (b) at the dominant harmonic \(k=3\) as a function of the finiteness factor \(f = \omega_1/(4\omega_c)\). The x-coefficients are identical for both sequences because the x-component of the trajectories is unchanged, while the y-coefficients highlight the decisive advantage of the phase adjustment in the finite-pulse regime. The original PulsePol experiments on NV centers in diamond~\cite{schwartz2018robust} correspond to the $f>10$ regime, whereas our experiments in Sec. \ref{6} probe PulsePol and Q-PulsePol in the $1<f<2$ regime.}
\label{fig:finiteness}
\end{figure}

We quantify pulse finiteness through the dimensionless ratio \(f = \omega_1/(4\omega_c)\). The finiteness factor ($f$) is chosen such that $f=1$ corresponds to the $\mu$w pulses collectively filling the full cycle $T_c$ (maximally finite). As \(f\) increases, the relative pulse duration decreases and the sequence approaches the ideal instantaneous-pulse limit. For PulsePol or Q-PulsePol to have a finiteness factor $f = 1$, the $\mu w$ power should be four-thirds of $\omega_{0n}$. To put this into perspective, the original PulsePol experiments on NV centers in diamond~\cite{schwartz2018robust}, performed at $\omega_1/(2\pi) \approx 50$~MHz with $^{13}$C nuclear Larmor frequency $\omega_{0n}/(2\pi) \approx 2$~MHz, correspond to $f \approx 18$. In contrast, our X-band experiments (in Sec. \ref{6}) with $\omega_1/(2\pi) = 38$~MHz and 21~MHz at $\omega_{0n}/(2\pi) \approx 15$~MHz ($^1$H at 0.35~T) correspond to $f \approx 1.9$ and $f \approx 1.05$.

Fig.~\ref{fig:finiteness} shows the magnitude of the Fourier coefficients \(|a_x^{(k)}|\) (a) and \(|a_y^{(k)}|\) (b) as a function of $f$ for both sequences for $k$=3. The \(|a_x^{(k)}|\) coefficient is essentially identical for the original PulsePol and Q-PulsePol across the entire range of $f$. This is expected, since the X-component of the interaction-frame trajectories remains unchanged by the central phase correction (as shown in Fig.~\ref{fig:finitepulse}b - black).

In contrast, the \(|a_y^{(k)}|\) coefficients exhibit a pronounced difference that becomes the critical differentiator. In the strongly finite-pulse regime (low \(f\), corresponding to lower microwave power), the original PulsePol suffers severe suppression of \(|a_y^{(k)}|\), while Q-PulsePol maintains a substantially higher amplitude. As microwave power increases (\(f\) becomes large), both sequences converge toward the ideal-pulse values. In fact, we can notice that Q-PulsePol has improved Fourier coefficients at lower $f$. This high efficiency of Q-PulsePol at reduced microwave powers is directly confirmed in the experimental results presented in Sec. \ref{6}.

The uni-modality and substantially higher Fourier coefficients of Q-PulsePol in the finite-pulse regime are expected to translate into markedly better polarization transfer efficiency and improved accumulation of bulk nuclear polarization through spin diffusion. To demonstrate these improvements directly, we performed numerical simulations of the DNP build-up (contact) curves.

\section{Bulk Nuclear Polarization} \label{5}

\begin{figure}[h!]
\centering
\includegraphics[width=0.925\linewidth]{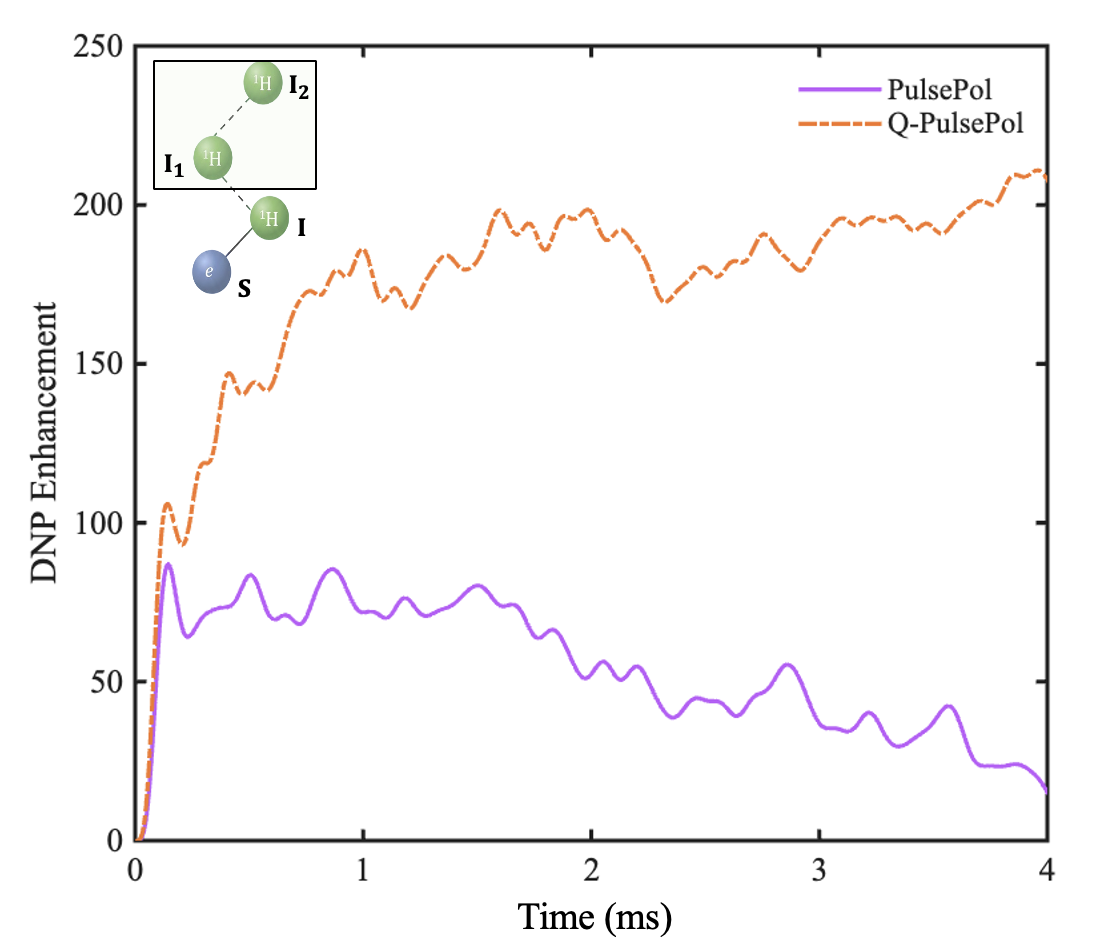}
\caption{\textbf{Simulated DNP buildup at Bulk Nuclear Spin Bath $\omega_1/(2\pi) = 38$ MHz.} Spin system with the bulk nuclear spin bath indicated by a box shaded in green and Simulated DNP buildup curve at $\omega_{0n} = 15$ MHz for PulsePol (purple) and Q-PulsePol (orange). The couplings are 37 kHz ${S}-{I}$, 20 kHz ${I}-{I}_1$, 10 kHz ${I}_1-{I}_2$. A PulsePol/Q-PulsePol cycle lasts $\approx$ 36 $\mu$s; 4 ms contact time is reached by repetition.}
\label{fig:contact}
\end{figure}

We numerically simulated DNP build-up (contact) curves for PulsePol and Q-PulsePol using SpinEvolution \cite{veshtort2006spinevolution} -- these are shown in Fig. \ref{fig:contact}. The spin system consists of an electron spin ($S$) that is dipolar (hyperfine) coupled to a nuclear spin ($I$) which is dipolar-coupled to multiple nuclear spins (here, ${I_1}$ and ${I_2}$); they form a chain where ${I_1}$ and ${I_2}$ mimic a spin bath (labeled in green). The couplings are purely dipolar (37 kHz ${S}-{I}$, 20 kHz ${I}-{I}_1$, 10 kHz ${I}_1-{I}_2$). Here, ${I}$ is only weakly coupled to ${S}$; this is deliberate, in order to represent a nucleus that will participate in efficient spin diffusion. Notably, the spin bath does not receive direct polarization from the electron spin, rather, it is polarized only indirectly via spin diffusion. The $\omega_{0n}/(2\pi)$ is 15\,MHz, while $\omega_1/(2\pi)$ is 38\,MHz. Each PulsePol or Q-PulsePol cycle lasts $\approx$ 36 $\mu$s; total contact times up to 4 ms are reached by repetition. During the simulation, the average of all hyperfine coupling orientations is considered.

Fig.~\ref{fig:contact} (purple) shows that original PulsePol hyperpolarizes the proximal nuclear spin ($I$) and thereby the nuclear spin bath ($I_1$ and $I_2$) at short contact times, however, this is followed by a steady decline at longer times. This occurs because the original PulsePol excites both DQ and ZQ resonances that destructively interfere, thereby reducing the net bulk polarization.

In contrast, Fig.~\ref{fig:contact} (orange) shows that Q-PulsePol produces smooth and robust build-up, with a steady polarization increase in the bulk bath ($I_1$ and $I_2$) for long contact times, even over 4 ms. Notably, we predict that the polarization fractions remain largely independent of $\mu$w nutation frequency because the restored Quadrature symmetry enforces a pure uni-modal DQ transfer pathway. Our experiments (Sec. \ref{6}) corroborate these simulation results where Q-PulsePol exhibits a monotonic build-up profile for both $\omega_1/(2\pi) =$ 21 MHz and 38 MHz, and notably demonstrates slightly better performance at the lower $\omega_1/(2\pi)$ frequencies, as anticipated in Sec. \ref{4}.

\section{Experimental Validation} \label{6}

\begin{figure*}
\centering
\includegraphics[width=0.98\textwidth]{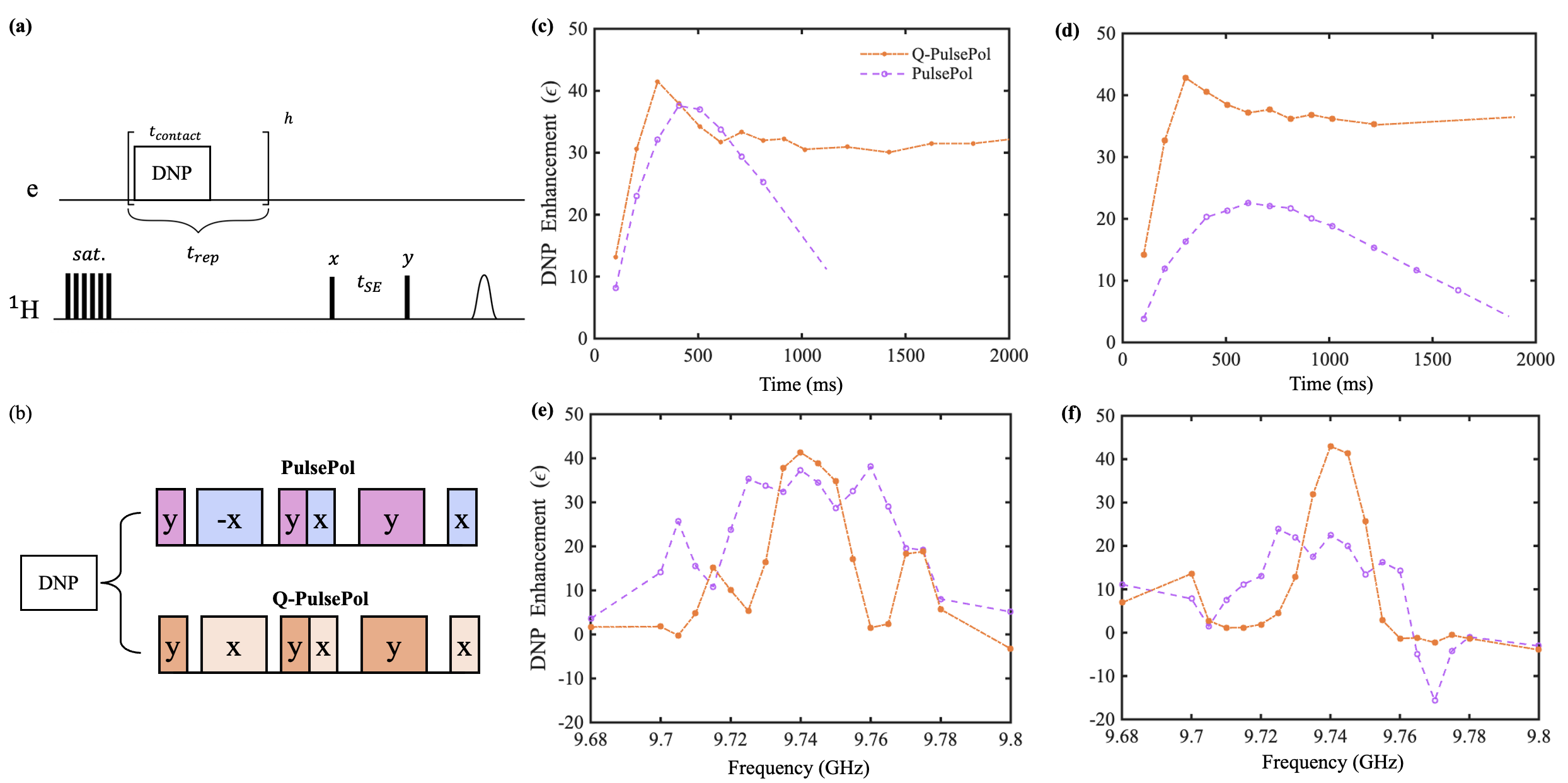}
\caption{\textbf{Experimental Results at 0.35 T.} (a) pulse-sequences used to measure the DNP enhancement, consisting of $^{1}H$ saturation train, followed by the DNP block where the total buildup time is defined by $T_{DNP}$ = $h\cross t_{rep}$, and ending on $^{1}H$ spin polarization readout with a solid-echo sequence. (b) DNP pulse-sequences used---PulsePol and Q-PulsePol. DNP build-up curves for PulsePol and Q-PulsePol at (c) $\omega_1/(2\pi)$ = 38 MHz and (d) $\omega_1/(2\pi)$ = 21 MHz. Offset profile of PulsePol and Q-PulsePol at (e) $\omega_1/(2\pi)$ = 38 MHz and (f) $\omega_1/(2\pi)$ = 21 MHz.}
\label{fig:exp}
\end{figure*}

All experiments were conducted on a home-built X-band pulsed EPR/DNP spectrometer (details listed in SI-4) and used a sample of 5 mM Trityl (OX063) in a H2O:D2O:Glycerol-d8 solution (1:3:6 by volume) at 80 K. Note that this sample at this temperature is a common system for pulsed-DNP studies using liquid nitrogen. At X-band (0.35\,T), the proton Larmor frequency is $\sim$15\,MHz, placing experimental nutation frequencies as high as 100\,MHz firmly in the finite-pulse regime, where pulse durations constitute a non-trivial fraction of $T_c$.

Pulsed DNP experiments were conducted using the pulse-sequence shown in Fig. \ref{fig:exp}(a, b) using initial saturation of $^{1}$H spin polarization with a set of S = 11 pulses of duration 1.32 $\mu$s, separated by $\tau_{sat}$ = 1 ms, and a solid-echo sequence $\pi/2-\tau-\pi/2$ with $\tau$ = 25 $\mu$s for readout. All RF pulses used an RF field strength of 230 kHz, corresponding to a $\pi/2$-pulse time of 1.10 $\mu s$. A 5 s overall pumping time with h = 2500 and a repetition time of $\tau_{rep}$ = 2 ms was used for all experiments. Experimental polarization enhancements (denoted $\epsilon$) are defined as the ratio between the DNP-enhanced signal intensity and the thermal equilibrium signal intensity.

The DNP enhancement profiles as a function of $\mu$w-offset for PulsePol and Q-PulsePol at $\omega_1/(2\pi)=38$\,MHz, where the pulses are less finite, are shown in Fig.~\ref{fig:exp}c. Both sequences reach nearly identical maximum DNP enhancement factors $\epsilon$, with PulsePol exhibiting a slightly broader offset bandwidth (see SI-3). However, when pulses are made more finite, $\omega_1/(2\pi)=21$\,MHz (Fig.~\ref{fig:exp}d), the enhancement of PulsePol falls to approximately half its high-power value while Q-PulsePol is slightly higher (see Sec. \ref{4}). This demonstrates the poor fidelity of standard PulsePol under finite $\mu$w pulses and the robustness of Q-PulsePol to finite-pulse durations. Moreover, it also underscores that the ideal-pulse approximation does not reliably predict pulsed-DNP performance at reduced nutation frequencies or higher magnetic fields.

Bulk build-up curves at $\omega_1/(2\pi)= 38$\,MHz (Fig.~\ref{fig:exp}e) show that bulk build-up for PulsePol reaches an initial rise followed by a steady decline. Clearly, even at this high nutation frequency, both DQ and ZQ pathways contribute to polarization transfer and, as discussed previously, suppress accumulation of bulk nuclear polarization via spin diffusion, despite the ZQ scaling factor being small. In contrast, Q-PulsePol displays a monotonic build-up profile. The Quadrature symmetry of the sequence enforces a pure uni-modal DQ pathway, enabling robust polarization transfer to the bulk nuclear spin bath through spin diffusion.

At $\omega_1/(2\pi)=21$\,MHz (Fig.~\ref{fig:exp}f), the ZQ scaling factor of PulsePol becomes comparable to the DQ contribution, resulting in substantially lower bulk polarization than observed at higher power. Q-PulsePol produces a build-up profile essentially identical to that at 38\,MHz, as its Quadrature symmetry preserves uni-modality independent of nutation frequency. These results establish that a uni-modal polarization-transfer pathway is required for efficient, long-time hyperpolarization of bulk nuclei via spin diffusion; finite-pulse-induced symmetry breaking in PulsePol activates detrimental competing resonances.

\section{Conclusions} \label{7}

We have shown that the ideal-pulse approximation fails to capture pulsed-Dynamic Nuclear Polarization (DNP) dynamics under realistic power constraints. Finite pulses explicitly break the toggling-frame symmetry \cite{carr1954effects, meiboom1958modified, waugh1968approach, cory1991new, ryan2010robust, equbal2017significance, tyler2023higher, joseph2025decoupling} in PulsePol triggering destructive Zero Quantum (ZQ) or Double Quantum (DQ) interference. Using Floquet Hamiltonian engineering, we isolated this mechanism by describing two fundamental symmetries: Quadrature and XY-Time-Reversal, and engineered Q-PulsePol, a minimal phase adjustment that restores symmetry and ensures robust, power-independent polarization transfer. Importantly, we have demonstrated that establishing a single-mode DNP transfer mechanism (either ZQ or DQ) is essential for generating net hyperpolarization of the bulk nuclei. We envision design of new schemes with lower required microwave power by applying the symmetry principles introduced here. Overall, our work bridges idealized quantum control with practical bounded-control experiments, significantly lowering hardware requirements for nano-scale NV sensing,  hyperpolarization of biomolcules, and quantum memory.

\subsection*{Acknowledgments} Contribution from AE was supported by Tamkeen under the NYU Abu Dhabi Research Institute grant CG006 and CG008. NCN acknowledges funding from the Villum Foundation Synergy programme (grant 50099), and the Novo Nordisk Foundation (NERD grant NNF22OC0076002). PKM acknowledge intramural funds at TIFR Hyderabad from the Department of Atomic Energy (DAE), India, under Project Identification Number RTI 4007.

\subsection*{Supplementary Material} 
Supplementary material, referred to as SI, includes additional equations, 2-D performance plots and experimental details.

\bibliographystyle{apsrev4-1}
\bibliography{pulsepol}   
\end{document}